\documentclass[preprint]{aastex}

 \usepackage{natbib}
 \usepackage{rotating}                                                                                     
\usepackage{graphicx}

\begin{document}                                                                     

\title{X-ray Detection of the Cluster Containing the Cepheid S Mus }


\author{Nancy Remage Evans\altaffilmark{1},
Ignazio Pillitteri\altaffilmark{1,2},
Scott Wolk\altaffilmark{1}, 
Edward Guinan\altaffilmark{3},
Scott Engle\altaffilmark{3}, 
Howard E. Bond,\altaffilmark{4}
Gail H. Schaefer,\altaffilmark{5} 
Margarita Karovska,\altaffilmark{1} 
Joseph DePasquale,\altaffilmark{1} 
and   
Evan Tingle\altaffilmark{1}  
}

\altaffiltext{1}
{Smithsonian Astrophysical Observatory,    
MS 4, 60 Garden St., Cambridge, MA 02138; nevans@cfa.harvard.edu}

\altaffiltext{2}
{INAF-Osservatorio Astronomico di Palermo, Piazza del Parlamento 1,
  90134  Palermo,  Italy}

\altaffiltext{3}{Department of Astronomy and Astrophysics, Villanova University, 800 Lancaster Ave. Villanova, PA 19085, USA}

\altaffiltext{4}
{Dept. of Astronomy and Astrophysics, Pennslyvania State University,
  University Park, PA 16802}

\altaffiltext{5}
{The CHARA Array of Georgia State University, Mount Wilson, CA, 91023 }




\begin{abstract}

The galactic Cepheid S Muscae has recently been added to the
important list of Cepheids linked to open clusters, in this case 
the sparse young  cluster ASCC 69. Low-mass
members of a young cluster are expected to have rapid rotation and
X-ray activity, making X-ray emission an excellent way to discriminate
them from old field stars. We have made an XMM-Newton observation
centered on S Mus and identified  (Table 1) a population of X-ray sources
whose near-IR 2MASS counterparts lie at locations in the J, (J-K)
color-magnitude diagram consistent with cluster membership at the
distance of S Mus.  Their median energy and X-ray luminosity are
consistent with young cluster members as distinct from field stars.
These strengthen the association of S Mus with the young
cluster, making it a potential Leavitt Law (Period-Luminosity
relation) calibrator.

\end{abstract}




\section{Introduction}


Galactic open clusters are an important means of calibrating
the absolute magnitude of Cepheids  (An, Terndrup, and Pinsonneault 2007;
Feast and Walker 1987; Turner and
Burke 2002). Recently  Anderson, Eyer, Mowlavi (2013, hereafter AEM)
made an all-sky survey of possible linkages between Cepheids and parent clusters
based on position, velocity, distance, abundance and age.  They found
a highly probable connection between the Cepheid S Mus and
the sparse cluster ASCC 69 = [KPR2005] 69 (Kharchenko, et al. 2005) 

The decrease in X-ray activity in low mass stars as they age and spin
down is well known (Pallavicini et al., 1981).  This means that X-ray activity provides an
excellent discriminant between young stars and the old field
population. Physical  companions of Cepheids must
be young, and hence  X-ray bright. 
We have used this approach to confirm possible resolved
companions of Cepheids (Evans, et al. 2013; Evans, et al. 2014 in
preparation) identified in a Hubble Space Telescope (HST) Wide Field
Camera 3 (WFC3) survey of 69 bright Cepheids.  In this study, we use
X-rays to identify low mass members of the cluster.  Identification of
low mass stars as cluster members is usually plagued by 
contamination of old field stars of similar colors, which limits the
value of this mass range in, for instance, determining the distance to
the cluster and studying the cluster population.  

\section{XMM Observation of S Mus}

 As part of the HST companion  program, we
observed the Cepheid S Mus with the XMM-Newton satellite (obsid: 0691030201, AO~11) on
January 5th 2013 for a total of 32.6 ksec with a 30' field of view.  
X-ray observations of the S Mus system itself will be discussed in a paper in preparation.
S~Mus lies outside the cluster radius of ASCC~69  as
listed in the catalog of Kharchenko, et al. (2005) but within two
cluster  radii.  However, AEM remark that
because the cluster is sparse, neither the center nor the radius is
well defined.

Because AEM found  S Mus to be a highly probable member of the cluster
ASCC 69, we investigated the X-ray sources in the entire XMM image to
see if there are any likely low mass cluster members. Source detection
was  performed in the same way as for the $\alpha$ Per cluster
(Pillitteri,  et al. 2013). 
  X-ray sources were then matched with 2MASS photometry
(Cutri, et al. 2003).  This filters the sources to eliminate background AGN,
which have  much fainter V magnitudes for a given X-ray flux than
stars.  Stars, on the other hand, at the distance of S Mus (789 pc, Evans, et
al. 2013) from the Leavitt period$-$luminosity law (Benedict et al., 2007)
 would be listed in the 2MASS catalog well into spectral class M. 
Only 2MASS sources with class AAA photometry have been included, which means the errors in
J and K are $\leq$ 0.10 mag.  

Fig 1 shows the results in a plot of  J vs. (J-K) for the
X-ray sources.  As indicated by the error bars, the main uncertainties 
are in the
colors. The isochrone from Siess \noindent\footnote{\sl
  http://www.astro.ulb.ac.be/~siess/Site/WWWTools} is overplotted 
for 30 Myr and solar abundance.   The discussion of
Bono, et al. (2005) shows that  S Mus (P = 9.65$^d$) has an age
between 30 and 50 Myr.  The isochrone (solid line) has been shifted
 to the distance of the Cepheid, with reddening
and A$_V$ of the Cepheid [E(B-V) = 0.21 mag; Evans et al 2013; E(J-K)
  = 0.10].   To illustrate the
expected location of cluster stars, a dashed line shows the isochrone
shifted by -0.8 mag to account for the binary sequence and +0.2 to
allow for errors in J-K.  A second dashed line
 has been shifted -0.2 in J-K.  These lines define four regions from
 left to right: A. background objects; B. a small region immediately
 behind/below the cluster sequence; C. the cluster sequence; and D. foreground objects.
The following number of stars are found in these regions:
A.  background: 8; B. below the cluster sequence: 2; C. the cluster sequence: 19; and
D. above/foreground: 3.  Thus, an over density of likely cluster members
is evident.   






\begin{figure}
 \includegraphics[width=5truein]{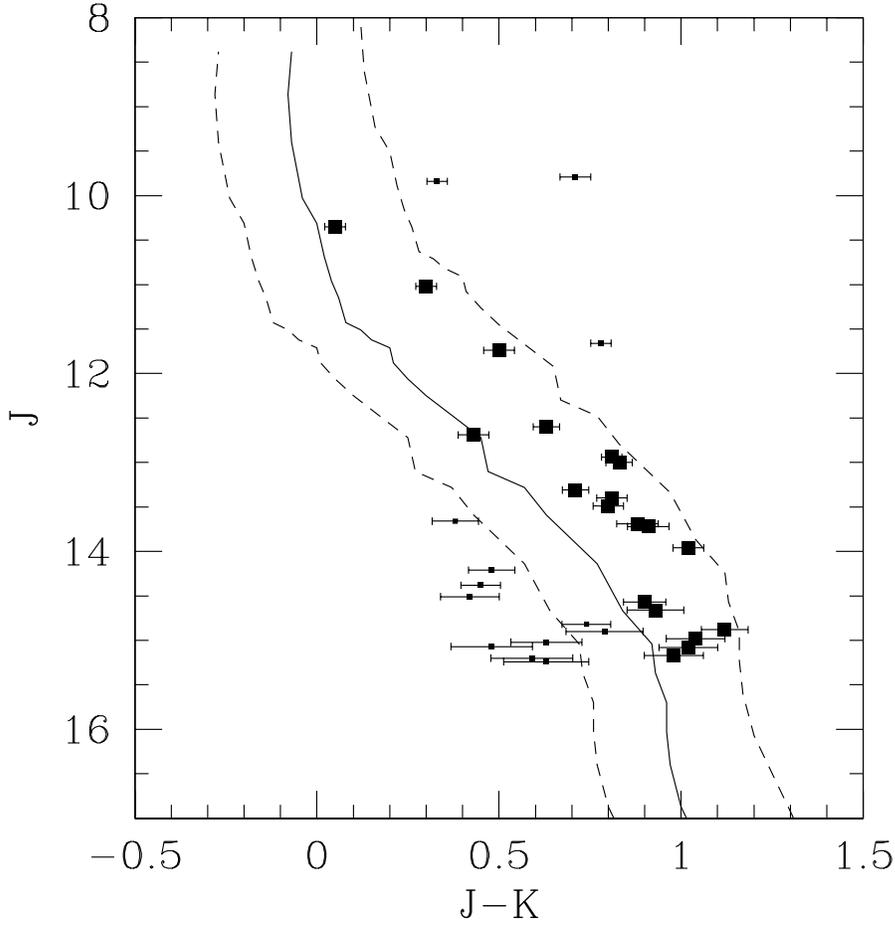}
\caption{The J--(J-K) color magnitude diagram for the X-ray sources in
  ASCC 69, the  cluster surrounding S Mus. Errors in J-K are shown;
  errors in J (Table 1) are smaller than the size of the symbols.
Sources identified as cluster members have larger symbols than
nonmembers. 
  The  Siess$^1$
isochrone for 30
  Myr  at the  distance and with the reddening of S Mus is shown
  (solid line). Cluster members are expected to lie below the upper
  dashed line which is 0.8
  mag  above (binaries) and 0.2 mag redder (errors) than the isochrone.
The lower dashed line is 0.2 mag bluer than the isochrone, below which
background stars would be found. 
 \label{fig1}}
\end{figure}


Table 1 lists the sources  considered to be young stars in the cluster region
in Fig. 1, that is stars which are 
X-ray sources and have the  J and K appropriate for a 30 Myr young cluster at the
distance of the Cepheid.  Cols 1 and 2 list the source number and the
source number in the original detection list respectively. 
Cols 3--5 list the 2MASS identification number and the right ascension
and  declination of the X-ray sources. Cols 6-9 list the 2MASS
photometry, J, H, K together with their errors, and J-K. 
The net count rate (net counts ksec$^{-1}$; Col 10) is from the detection
routine. The median X-ray energy in keV is  in Col 11.
We have used {\it PIMMS} to derive a conversion factor from
the count rate to flux for a 1-T APEC spectrum with kT = 1.1 keV and
N$_H$ = 1.0 x 10$^{21}$, appropriate for the E(B-V).  Luminosities
have been computed for the fluxes for the distance of the Cepheid (789
pc).  These are provided in cols 12 and 13 respectively. The final
column (14) provides the separation between the X-ray source and the
2MASS source.    

We have examined the X-ray parameters of the probable cluster members
to see if they are consistent
with stellar coronal sources.  Median energies are typically in the
range of 1 to 2 keV, which is reasonable for young stellar sources   
(e.g. Feigelson, et al. 2005).  Stars from
Figure 1 which fall outside the cluster band have similar median
energies, indicating that they may  also be stars, but not spatially
coincident with the cluster.  A typical Median Absolute Deviation
(MAD) is 0.6.  Only 4 of the sources have fewer than 30 counts.  For
the others, using Gaussian statistics and a typical source count of 50
counts, the uncertainty in the median energy is estimated to be 0.1 keV.  

In order to have a consistent estimate of the character of the
sources, 
we have used median energies  to provide information about all
the sources which are probable low-mass cluster members (Table 1).
Two of the sources have enough counts to fit a spectrum, \#60 (273
counts), and \#72 (441 counts).  The next strongest source has only
half as many counts (\#7 137 counts) and all the rest of the sources
have fewer than 100 counts.  For the two strong sources (\#60 and
\#72) the results of the spectral fitting are as follows. 
 For source 60, the N$_H$, kT, and log Flux
(unabsorbed) are 1.3 $\times$ 10$^{21}$, 0.93 keV, and -13.56  (ergs cm$^{-2}$
s$^{-1}$). For source 72, they are  1.8 $\times$ 10$^{21}$, 1.43 keV, and -13.16
(ergs cm$^{-2}$ s$^{-1}$).  These  are similar to the N$_H$ derived
from E(B-V), and the values for kT and log Flux  listed in Table
1. These values would be expected for stellar coronal sources rather
than background AGN.

The width of the probable cluster band in Fig. 1 implies a significant
population of binaries.  In addition we note that for  M25, the cluster
which contains the Cepheid U Sgr (Majaess, et al. 2013), the main
sequence in the J--(J-H) CMD is markedly wider than that in the
V--(B-V) CMD.  There may also be a significant width to the J--(J-K) CMD
for the $\alpha$ Per cluster, which is similar in age to Cepheids
(Pillitteri, et al. 2013).  This points to complexity in interpreting
the cluster data, which is affected by the accruacy of photometry, the
population of binaries, a spread in ages or even possibly the
existance of debris disks (which would be more prominent in infrared
colors).  It would be premature to draw conclusions from the data in
Fig. 1 on this topic, but will be discussed in a further paper.

The exposure is  comparatively shallow for sources $\sim800$~pc away, so
only the brightest part of the X-ray luminosity distribution will be
sampled.  The X-ray luminosities (Col 13) are reasonable for the
brightest  stellar sources at the age of the cluster (eg. Pillitteri
et al, 2013; Preibisch and Feigelson 2005).  We also compare the
luminosities of the cluster stars with the putative background stars.  
The luminosities of the stars below the lower dashed line in Fig. 1
(region A using the distance of the cluster, which is, of course, an
underestimate)  typically appear to 
 have $\log L_X$ $\leq$ 30 ergs sec$^{-1}$ while those of the
cluster members (Col 13) spread to significantly brighter
luminosities.  Again, this is appropriate for stars actually more distant than
the cluster.   

Future work includes a comparison of Fig. 1 with the
low-mass  members of other clusters associated with Cepheids from both new
observations and archival data,  particularly pertaining to the width of the
main sequence band.  
Included in this more extensive study will be 
the data in the current study and for M25, as well as an 
additional cluster observation that has been approved for XMM, permitting a
discussion with a wider scope.  


\section{Conclusions}

An XMM-Newton observation centered on the Cepheid S Mus identifies 
 a concentration of 19 X-ray sources with 2MASS magnitudes
appropriate for a 30 Myr cluster at the distance of the Cepheid.
  These
 are  low mass stars which are likely cluster members, supporting the
 identity of the sparse cluster ASCC 69.  The paper  demonstrates 
 the value of using X-ray observations to identify young X-ray active
 low-mass cluster
 candidates from a large number of older stars in the
 field. Confirmation of the cluster strengthens the association of S
 Mus with the cluster.

\acknowledgments

We thank an anonymous referee for comments which improved the
presentation of the paper.  
 Support for this work was also provided  from the Chandra X-ray Center NASA 
Contract NAS8-03060 and  by HST grant GO-12215.01-A.  
Vizier and  SIMBAD were used in the preparation of this study.

\clearpage


\begin{sidewaystable}
    \centering
    \caption{List of cluster members defined as the 2MASS objects with X-ray detection falling in the region of Fig. 1 identified with the cluster
sequence.}
\resizebox{0.9\textheight}{!}{
\begin{tabular}{llcllcccclllll}
\hline  
\hline  
N & Star & 2MASS &  RA & Dec &  J  $\pm$ E$_J$ &  H  $\pm$ E$_H$ &  K  $\pm$
E$_K$ & J-K & Ct Rt & Med E  & log F$_X$ & log L$_X$ & Sep \\
 &   &   & deg (J2000) &  deg (J2000) &  mag  $\pm$ mag &  mag  $\pm$ mag &  mag $\pm$ mag &  mag & ks$^{-1}$ & keV &  ergs cm$^{-2}$s$^{-1}$ &
ergs s$^{-1}$ & \arcsec \\
 1 & 2 & 3 & 4 & 5 & 6 & 7 & 8 & 9 & 10 & 11 & 12 & 13 & 14 \\
\hline  
1 & 7 & 12133153-7016575 & 183.38107 & -70.28242 & 11.74$\pm$0.03 & 11.33$\pm$0.03 & 11.24$\pm$0.03 & 0.51 & 3.2 & 1.02 & -13.61 & 30.26 & 1.0 \\ 
  2 & 11 & 12140653-7016261 & 183.52871 & -70.27355 & 14.66$\pm$0.05 & 13.92$\pm$0.04 & 13.73$\pm$0.06 & 0.92 & 1.7 & 1.46 & -13.88 & 30.00 & 2.3 \\ 
  3 & 18 & 12121521-7013476 & 183.06439 & -70.22959 & 11.02$\pm$0.02 & 10.75$\pm$0.02 & 10.72$\pm$0.02 & 0.3 & 0.9 & 0.98 & -14.17 & 29.70 & 1.7 \\ 
  4 & 33 & 12111525-7009518 & 182.81650 & -70.16427 & 13.00$\pm$0.02 & 12.32$\pm$0.02 & 12.17$\pm$0.03 & 0.83 & 0.4 & 1.48 & -14.55 & 29.33 & 3.6 \\ 
  5 & 36 & 12105447-7009296 & 182.72909 & -70.15813 & 13.69$\pm$0.04 & 13.05$\pm$0.04 & 12.81$\pm$0.04 & 0.88 & 1.8 & 1.01 & -13.87 & 30.01 & 2.6 \\ 
  6 & 37 & 12144088-7009232 & 183.67194 & -70.15599 & 14.88$\pm$0.04 & 14.22$\pm$0.04 & 13.76$\pm$0.05 & 1.12 & 0.7 & 1.14 & -14.24 & 29.63 & 2.6 \\ 
  7 & 38 & 12123024-7009130 & 183.12328 & -70.15525 & 12.94$\pm$0.02 & 12.25$\pm$0.02 & 12.13$\pm$0.02 & 0.82 & 0.9 & 2.49 & -14.15 & 29.72 & 6.7 \\ 
  8 & 45 & 12133084-7008364 & 183.37656 & -70.14260 & 12.69$\pm$0.03 & 12.35$\pm$0.03 & 12.26$\pm$0.03 & 0.44 & 0.5 & 0.98 & -14.41 & 29.46 & 3.9 \\ 
  9 & 46 & 12133429-7008332 & 183.39420 & -70.14211 & 15.08$\pm$0.04 & 14.31$\pm$0.04 & 14.06$\pm$0.07 & 1.02 & 0.3 & 1.01 & -14.62 & 29.25 & 2.3 \\ 
  10 & 47 & 12104829-7008261 & 182.70175 & -70.14057 & 14.98$\pm$0.04 & 14.32$\pm$0.04 & 13.94$\pm$0.07 & 1.04 & 0.5 & 1.36 & -14.40 & 29.48 & 0.7 \\ 
  11 & 51 & 12143476-7006537 & 183.64800 & -70.11479 & 13.96$\pm$0.03 & 13.27$\pm$0.03 & 12.94$\pm$0.03 & 1.02 & 2.5 & 1.15 & -13.71 & 30.16 & 3.9 \\ 
  12 & 52 & 12101845-7006448 & 182.57796 & -70.11176 & 10.35$\pm$0.02 & 10.34$\pm$0.02 & 10.30$\pm$0.02 & 0.05 & 1.0 & 1.13 & -14.10 & 29.77 & 2.9 \\ 
  13 & 60 & 12122464-7005448 & 183.10372 & -70.09562 & 13.72$\pm$0.04 & 13.12$\pm$0.04 & 12.81$\pm$0.04 & 0.91 & 4.5 & 1.10 & -13.46 & 30.41 & 1.4 \\ 
  14 & 71 & 12124049-7004243 & 183.16720 & -70.07364 & 15.17$\pm$0.04 & 14.50$\pm$0.06 & 14.19$\pm$0.07 & 0.98 & 0.8 & 1.29 & -14.21 & 29.67 & 2.0 \\ 
  15 & 72 & 12124932-7004087 & 183.20683 & -70.06896 & 13.49$\pm$0.03 & 12.79$\pm$0.02 & 12.69$\pm$0.03 & 0.80 & 7.6 & 1.11 & -13.23 & 30.64 & 1.7 \\ 
  16 & 76 & 12120080-7003457 & 183.00435 & -70.06203 & 14.57$\pm$0.03 & 13.9$\pm$0.03 & 13.67$\pm$0.05 & 0.90 & 2.0 & 1.00 & -13.80 & 30.07 & 2.8 \\ 
  17 & 77 & 12150330-7003181 & 183.76350 & -70.05444 & 13.31$\pm$0.03 & 12.77$\pm$0.02 & 12.60$\pm$0.02 & 0.71 & 2.2 & 1.05 & -13.76 & 30.11 & 2.2 \\ 
  18 & 88 & 12105702-6959592 & 182.73689 & -69.99788 & 12.60$\pm$0.03 & 12.11$\pm$0.03 & 11.97$\pm$0.02 & 0.63 & 1.8 & 2.07 & -13.86 & 30.01 & 7.0 \\ 
  19 & 94 & 12133206-6955321 & 183.38331 & -69.92509 & 13.40$\pm$0.03 & 12.71$\pm$0.03 & 12.59$\pm$0.03 & 0.81 & 5.4 & 1.40 & -13.38 & 30.49 & 1.9 \\ 
\hline
    \end{tabular}
}
\end{sidewaystable}

\end{document}